\documentstyle[preprint,aps]{revtex}

\begin{document}
\author{J. Twamley%
\thanks{New address: 
     Department of Physics, Blackett Laboratory,
     Imperial College of Science, Technology and Medicine,
     Exhibition Road,
     London SW7 2BZ,   U.K.}}
\title{Bures and Statistical Distance for Squeezed Thermal States}
\date{14 March 1996}
\maketitle

\begin{abstract}
We compute the Bures distance between two thermal squeezed states and deduce
the Statistical Distance metric. By computing the curvature of this metric
we can identify regions of parameter space most sensitive to changes in
these parameters and thus lead to optimum detection statistics.
\end{abstract}

\section{Introduction}

There has been an expanding interest recently in the geometry of quantum
state space \cite{INTEREST,Intro}. Although some general features of the
geometry are known, little has been learned concerning the details of the
geometry of high dimensional pure and impure states. The initial discovery
of a geometric phase by Berry \cite{BERRY1} was interpreted by Simon \cite
{SIMON} as the holonomy transformation in parallel transporting the
adiabatic eigenstate in parameter space \cite{SEN}. Since then the concept
of the geometric phase has been broadened to cope with non-adiabatic,
non-cyclic and non-unitary evolutions \cite{BERRY2}. However, although the
formal understanding of the geometric phase has progressed, knowledge of the
underlying geometry described by this phase has not. This is mainly due to
computational difficulties in calculating metric tensors. These metric
tensors are functionals of infinite dimensional density matrix operators. In
this paper we calculate the metric and curvature of the parameter space of
squeezed thermal quantum states. Using a basic understanding of the quantum
metric from a statistical inference viewpoint we identify regions of
parameter space which yield large changes in the quantum state making its
determination easier in such parameter regime. In section II we review the
work done to date in uncovering the geometry of a quantum state and
introduce the Bures metric. The main results of this paper is the method of
calculation and final results in section III. We have tried to keep this
short as short as possible while including enough detail for the reader to
reproduce their own calculation.

\section{Review}

In this section we review the basics of the geometry of quantum states and
introduce concepts associated with the natural extension of the Fubini-Study
metric to impure density matrices and in particular, the Bures metric.

For pure quantum states the geometry is $CP^n$ and is essentially the
geometry of the horizontal section of the fibre bundle over the space of
pure states with a fibre group $U(1)$. The connection defining this section
is ``natural'' in that the resulting metric and distance functions are
invariant under a global change in phase of the states involved. More
precisely, the expectation value of any operator in the state $|\psi \rangle 
$ is unchanged under the action $|\psi \rangle \rightarrow e^{i\theta }|\psi
\rangle $. We can thus split up the space of pure states into conjugacy
classes under the $U(1)$ action and denote the class formed from the state $%
|\psi \rangle $, the {\em ray} at $|\psi \rangle $. We can define a distance
between two rays as the smallest transition probability between any two
elements in the separate rays ie. $D_{FS}^2=\inf ||\,|\psi _1\rangle
-e^{i\theta }|\psi _2\rangle ||^2$. Extreamising over the relative phase $%
\theta $ we obtain the well known Fubini-Study distance for pure states $%
D_{FS}^2=2(1-|\langle \psi _1|\psi _2\rangle |^2)$. One can show that the
geometry is K\"{a}hler and the metric, $ds_{FS}^2$, is the Hessian of a
suitable K\"{a}hler potential \cite{KAHLER}. This Riemannian metric arises
in calculations of Berry's phase and generalisations thereof \cite{BERRY2}
and in aspects of quantum distinguishability \cite{BRAUNSTEINCAVES}. The
geometry of {\em impure} quantum states has received little attention. A
Riemannian metric for classical probability distributions was obtained
independently by Wootters \cite{WOOTTERS} and Campbell \cite{CAMPBELL}. A
transition amplitude between two impure quantum states was discovered by
Bures \cite{BURES}. This amplitude and the related metric has been studied
at length with regard to geometric phase by Uhlmann \cite{UHLMANN}. The same
metric has also been obtained in other work relating to the optimal
statistical distinguishability between two quantum states \cite
{BRAUNSTEINCAVES}.

Although some formal work has been done on the geometry of impure quantum
states few concrete results concerning the details of the metric have been
found. This is due to the technical difficulties in computing the Bures (or
Statistical Distance) metric. Before giving the formula we will outline
briefly the origins of this metric following Uhlmann. The derivation follows
the above argument for the Fubini-Study metric on pure states. Beginning
with an impure state $\rho$ one {\em purifies} this state by enlarging the
Hilbert space into a Hilbert-Schmidt space through the ``square-root" of $%
\rho$, i.e. ${\cal H}\rightarrow {\cal H}^{ext}\equiv {\cal H}\otimes{\cal H}%
^*$ where $\rho\equiv WW^*$, $W$ is pure in ${\cal H}^{ext}$ and where ${\rm %
Tr}\, WW^*<\infty$. The ``square-root" $W$, of $\rho$ is defined up to right
multiplication by an arbitrary unitary operator $V$. We again have a fibre
bundle structure with base space $\sqrt{\rho}$ and fibre $R_{V}$, where $R_V$
is right multiplication by $V$. The natural distance in ${\cal H}^{ext}$ is
the Hilbert-Schmidt metric $d^2(W_1,W_2)\equiv {\rm Tr}\,(W_1-W_2)(W_1-W_2)^*
$. This gives a natural connection on the bundle and one can again define a
distance between two fibers to be the smallest Hilbert-Schmidt distance
between elements of the fibres. The solution to the extreamisation of $%
W(\lambda)$ where $\lambda$ is an affine parameter, is $\dot{W}=GW$ where $%
G=G^*$ and ${}^.\equiv d/d\lambda$. The induced metric on the horizontal
section is just 
\begin{equation}
(\dot{W},\dot{W})_{HS}=(GW,GW)_{HS}= {\rm Tr}\,G^2\rho=\frac{1}{2}{\rm Tr}\,G%
\dot{\rho}\;\;.  \label{123}
\end{equation}
The extreamised $W(\lambda)$ is parallel transported with respect to the
natural connection and gives rise to an evolution for $\rho(\lambda)$ in $%
{\cal H}$ which obeys 
\begin{equation}
\dot{\rho}=G\rho+\rho G\;\;.  \label{124}
\end{equation}
The Bures distance which results from the extreamisation can be written as 
\begin{equation}
D_B^2(\rho_1,\rho_2)= 2\left[1-{\rm Tr}\,\sqrt{\rho_1^{1/2}\,\rho_2\,%
\rho_1^{1/2}}\right] \;\;,  \label{125}
\end{equation}
while the infinitesimal Riemannian metric resulting from this distance is 
\begin{equation}
ds^2_B\equiv{\rm Tr}\,G^2\rho=\frac{1}{2}{\rm Tr}\,G d\rho\;\;,  \label{126}
\end{equation}
where 
\begin{equation}
d\rho=G\rho+\rho G\;\;.  \label{127}
\end{equation}
This metric is also known as the Statistical Distance metric and is
symmetric in $\rho_1,\;\rho_2$ \cite{BRAUNSTEINCAVES}.

Investigation into the detailed structure of the Bures distance has been
hampered by the complicated square-root factors in (\ref{125}). The distance
and metric have been calculated for the spin-1/2 system \cite{HUBNER1} and
the spin-1 system \cite{HUBNER,DITTMANN}. It was found that the geometry of
spin-1/2 state space was of constant curvature. However, the geometry of the
spin-1 state space possessed a non-constant curvature. It was further proved
in \cite{DITTMANN1} that the geometry of state space for spin-$n$ is not of
constant curvature and not even locally symmetric for $2n+1\geq 3$. To
directly solve for the Riemannian metric (\ref{126}) one must solve the
matrix Lyapunov equation (\ref{127}). For quantum systems possessing a
finite dimensional representation the method of annihilating polynomials can
be used to solve the Lyapunov equation \cite{LYAPUNOV}. This becomes
prohibitive for $n>3$ and results in non-unique expressions with respect to
the parametrisations chosen for $\delta\rho$. Other solution methods are
available but are again difficult to compute i.e. recursive solutions.

In the following we will first calculate the Bures distance between two
undisplaced thermal squeezed states and from this derive the associated 
Riemannian metric. Calculating the curvature of this metric we find the
space is not of constant  curvature and can interpret this curvature as a
measure of optimal quantum distinguishability between the states.

\section{Bures Distance}

From the work of Bures and Uhlmann \cite{BURES,UHLMANN} the transition
amplitude between two quantum states may be written as 
\begin{equation}
D_{{\rm B}}^2(\rho_1,\rho_2)= 2\left[1-{\rm Tr}\sqrt{\rho_1^{1/2}\rho_2%
\rho_{1}^{1/2}}\,\right]\;\;.  \label{1}
\end{equation}
Due to the complexities of computing the trace, studies of this transition
amplitude have concentrated only on finite dimensional examples with
concrete results for dimension $2$ \cite{HUBNER} and $3$ \cite{DITTMANN}. In
the following we will compute the transition amplitude between two thermal
squeezed states with density matrices parametrised in the form: 
\begin{equation}
\rho(\beta,r,\theta)=ZS(r,\theta)T(\beta)S^\dagger(r,\theta)\;\;,  \label{2}
\end{equation}
where 
\begin{eqnarray}
S(r,\theta) &=&\exp\left(\zeta K_+-\zeta^* K_-\right)\;\;,  \nonumber \\
T(\beta)&=&\exp\left(-\beta K_0\right)\;\;,  \nonumber  \label{3} \\
\zeta&=&re^{i\theta}\;\;,  \nonumber
\end{eqnarray}
and 
\begin{eqnarray}
K_+=\frac{1}{2}{a^\dagger}^{2}\;,\; \;& K_ -=\frac{1}{2}a^2 \;,\;\; & K_0= 
\frac{1}{2}(a^\dagger a+\frac{1}{2})\;\;,  \nonumber
\end{eqnarray}
\begin{eqnarray}
\left[ K_0,K_\pm\right]=\pm K_\pm \;,\;\; & \left[K_-,K_+\right]=2K_0\;\;. 
\nonumber
\end{eqnarray}
Here $S(r,\theta)$ is the one photon squeeze operator, $a$ is the single
mode annihilation operator, $Z$ is chosen so that ${\rm Tr}(\rho)=1$, and $%
(K_0,K_\pm)$ are the generators of the SU(1,1) group. Equation (\ref{2})
thus represents an undisplaced squeezed thermal state. We have written the
density matrix in the Schur form $\rho=UTU^\dagger$ where $U$ is unitary and 
$T$ is diagonal in the eigenbasis of $K_0$. This decomposition is relatively
straightforward in the case of gaussian $\rho$ \cite{KRUGER}. However, we
find that the following arguments do not seem to hold if we expand the
states considered to include displaced thermal squeezed states. We are thus
restricted to density matrices continuously paramterised by three variables $%
(\beta, r,\theta)$.

In the following this Shur factorisation will play a central role. With this
factorisation we can easily define the square root of a positive operator. A
possible alternative method is to represent the squeeze and thermal
operators as 2$\times$2 matrix represntations of SU(1,1). However, it is
unclear to the author at this time how one can consistently define the
square root in this representation without again forming the Shur
decomposition of the matrix representation. We now outline how the Shur
decomposition yields the square root of a positive operator $\rho$.

Through an insertion of unity, the Schur factorisation of $\rho$ yields a
diagonal representation of the state. Choosing orthogonal eigenstates $%
|\lambda_i\rangle$, such that $T(\beta)|\lambda_i\rangle=\lambda_i(\beta)|%
\lambda_i\rangle$ and $Z\sum_{i}\lambda_i =1$, we can insert the resolution
of unity $\sum_i |\lambda_i\rangle\langle\lambda_i|\equiv 1\!\!1$ into (\ref
{2}) to get 
\begin{eqnarray}
\rho&&=ZUTU^\dagger=ZUT1\!\!1U^\dagger  \nonumber \\
&&=Z\sum_{i}UT|\lambda_i\rangle\langle\lambda_i|U^\dagger=
Z\sum_{i}\lambda_iU|\lambda_i\rangle\langle\lambda_i|U^\dagger  \nonumber \\
&&=Z\sum_{i}\lambda_i|u_i\rangle\langle u_i|\;\;,  \nonumber  \label{4}
\end{eqnarray}
where $|u_i\rangle\equiv U|\lambda_i\rangle$, $\langle u_i|u_j\rangle=
\delta_{ij}$ and ${\rm Tr}(\rho)=1$. Thus we have diagonalised $\rho$ over a
complete orthonormal set of states with corresponding probabilities $P_i=
Z\lambda_i$. It is now an easy matter to find $\rho^{1/2}$: 
\begin{equation}
\rho^{1/2}=W=\sum_i P_i^{1/2}|u_i\rangle\langle u_i|V\;\;,  \label{5}
\end{equation}
where $V$ is an arbitrary unitary operator and $\rho=WW^\dagger$.
Essentially, $V$ encodes the ambiguity in taking the square root of an
infinite dimensional operator. To evaluate the trace in (\ref{1}) we need
only compute $\rho_1^{1/2}$. From the alternate definition of the Bures
Distance 
\begin{equation}
D^2_B=(\sqrt{\rho_1},\sqrt{\rho_2})_{HS}= \inf\;{\rm Tr}(W_1-W_2)(W_1-W_2)^%
\dagger\;\;,  \nonumber  \label{6}
\end{equation}
where $\rho_1=W_1W_1^\dagger$, $\rho_2=W_2W_2^\dagger$, we see that the
definition (\ref{1}) is invariant under the transformation $W_i \rightarrow
W_i \tilde{V}$ where $\tilde{V}\tilde{V}^\dagger=1$. By right multiplying $%
W_1$ and $W_2$ by $V_1^ \dagger$ in (\ref{1}) we can shift away the $V_1$
dependence of $\sqrt\rho_1$ to get 
\begin{equation}
\rho_1^{1/2}=\sum_{i} P_i|u_i\rangle\langle u_i|\;\;,  \nonumber  \label{7}
\end{equation}
Since we need not calculate $\rho_2^{1/2}$ we can ignore the unitary $%
V_2V_1^\dagger$ appearing in $\rho_2^{1/2}$.

Let us now summarise the manipulations needed to calculate the complicated
trace factor in (\ref{1}). Given $\rho_1$ we can now compute $\sqrt\rho_1$
taking the positive section for the square root. Using $\rho_2$ we form the
operator $A=\rho_1^{1/2}\rho_2\,\rho_1^{1/2}$ and, using the
Baker-Campbell-Hausdorff identities, we rearrange $A$ into Schur form 
\begin{equation}
A=U_AT_AU_A^\dagger\;\;.  \label{8}
\end{equation}
In this representation we can easily compute the square root, again taking
the positive section. All that remains is to take the trace. The
rearrangement of $A$ into Schur form is not trivial. The operators $U_A$ and 
$T_A$ can be found only in the case of undisplaced squeezed thermal states.
For displaced states a Schur resolution was not found through BCH
disentangling. It may be that in the case of displaced states the positive
section for the square-root is not the correct ansatz and the more general
form (\ref{5}) is needed. We will not address this here and will only
consider states of the form (\ref{2}) which do result in Schur
decompositions for $A$.

We now calculate ${\rm Tr}\sqrt{\rho_1^{1/2}\rho_2\rho_1^{1/2}}$ where 
\begin{equation}
\rho_i=Z_iS(r_i,\theta_i)\exp(-\beta_i K_0) S^\dagger(r_i,\theta_i)\;\;, 
\nonumber  \label{9}
\end{equation}
with normalisation ${\rm Tr}(\rho)=1$ or, $Z=2\sinh \beta_i/4$. Writing $%
T_i=\exp(-\beta_i K_0)$ we must rearrange $\rho_1^{1/2}\rho_2\rho_1^{1/2}$
to have the Schur form: 
\begin{eqnarray}
\rho_1^{1/2}\rho_2\,\rho_1^{1/2}& =&Z_1Z_2S_1T_1^{1/2}S_1^\dagger
S_2T_2S_2^\dagger S_1T_1^{1/2\dagger} S_1^\dagger  \nonumber \\
&=& Z_1Z_2S_1S_3T_3S_3^\dagger S_1^\dagger\equiv A\;\;.  \label{10}
\end{eqnarray}
Taking the positive square root and trace gives: 
\begin{eqnarray}
{\rm Tr}\sqrt{A}&=&\sqrt{Z_1Z_2}\,{\rm Tr}\, e^{-\beta_3/2 K_0}\;\;, 
\nonumber \\
&=& \frac{\sqrt{\sinh \beta_1/4 \sinh \beta_2 /4}}{\sinh \beta_3/8}\;\;.
\label{11}
\end{eqnarray}

We must now use Baker-Campbell-Hausdorff identities to express $\beta_3$ in
terms of $(\beta_1,\beta_2,r_1,r_2,\theta_1,\theta_2)$. This first step is
to collapse the product $S_1^\dagger S_2$ in (\ref{10}) into a single
squeeze operator. This is accomplished through the identity \cite{SCHMUACKER}
\begin{equation}
S^\dagger (r_1,\theta_1)S(r_2,\theta_2)=e^{-i\phi}\bar{S}(\bar{r}, \bar{%
\theta}-\phi)\bar{R}(\phi)\;\;,  \nonumber  \label{12}
\end{equation}
where $R(\phi)=e^{i\phi K_0}$, is the rotation operator while the parameters 
$(r_i,\theta_i)$ are related to $(\bar{r},\bar{\theta},\phi)$ through 
\begin{equation}
C_{\bar{r}\bar{\theta}}e^{i\phi\sigma_3}=C_{r_2\theta_2}C_{r_1\theta_1}\;\;,
\nonumber  \label{13}
\end{equation}
where 
\begin{equation}
C_{r\theta}=\left[ 
\begin{array}{cc}
\cosh r & e^{2i\theta}\sinh r \\ 
e^{-2i\theta}\sinh r & \cosh r
\end{array}
\right]\;\;,  \label{12.5}
\end{equation}
and $\sigma_3$ is the third Pauli matrix. Collapsing this product in (\ref
{10}) gives 
\begin{equation}
\rho_1^{1/2}\rho_2\rho_1^{1/2}=Z_1Z_2S_1T_1^{1/2}\bar{S}\bar{R}T_2\bar{R}^
\dagger\bar{S}^\dagger T_1^{1/2}S_1^\dagger\;\;.  \nonumber
\end{equation}
However, $\bar{R}T_2\bar{R}^\dagger=T_2$ since $\bar{R}$, $T_2$ commute. The
factor $\exp (i\phi)$ cancels since it is a scalar. We thus have 
\begin{equation}
\rho_1^{1/2}\rho_2\rho_1^{1/2}=Z_1Z_2S_1T_1^{1/2}\bar{S}T_2\bar{S}^\dagger
T_1^{1/2}S_1^\dagger\;\;.\;  \nonumber  \label{13.2}
\end{equation}
We now must rearrange the product $T_1^{1/2}\bar{S}T_2\bar{S}^\dagger
T_1^{1/2}S_1^\dagger$ into Schur form. To do this we use
Baker-Campbell-Hausdorff disentangling. The particular method we use was
outlined in \cite{GILMORE}. Using the faithful group representation of
SU(1,1) where 
\begin{equation}
K_+=\left[ 
\begin{array}{ll}
0 & 1 \\ 
0 & 0
\end{array}
\right] \;\; K_-=\left[ 
\begin{array}{ll}
0 & 0 \\ 
-1 & 0
\end{array}
\right]\;\; K_0=\frac{1}{2}\left[
\begin{array}{ll}
1 & 0 \\ 
0 & -1
\end{array}
\right]\;\;,  \nonumber
\end{equation}
we can express the operator $\bar{S}$ as 
\begin{equation}
\bar{S}=\exp (\bar{\zeta} K_+-\bar{\zeta}^* K_- )=\left[
\begin{array}{ll}
\cosh \bar{\gamma} & e^{i\bar{\theta}}\sinh \bar{\gamma} \\ 
e^{-i\bar{\theta}}\sinh \bar{\gamma} & \cosh \bar{\gamma}
\end{array}
\right]\;\;,  \nonumber
\end{equation}
where $\bar{\gamma}=\sqrt{\bar{\zeta}\bar{\zeta}^*}$ and $\bar{\theta}= 
\sqrt{\bar{\zeta}/\bar{\zeta}^*}$. The operator $T_i$ is represented as 
\begin{equation}
T_i=\left[
\begin{array}{ll}
e^{-\beta_i/2} & 0 \\ 
0 & e^{\beta_i/2}
\end{array}
\right]\;\;.  \nonumber
\end{equation}
To re-express the product $T_1\bar{S}T_2\bar{S}^\dagger T_1$ as $ST_3
S^\dagger$ we represent each operator as a 2x2 matrix, multiply and compare
the resulting entries to obtain 
\begin{eqnarray}
&&\cosh \beta_3/2=  \nonumber \\
&&\cosh ^2\bar{\gamma} \cosh((\beta_1+\beta_2)/2)- \sinh ^2\bar{\gamma}%
\cosh((\beta_1-\beta_2)/2) \;\; .  \nonumber  \label{14}
\end{eqnarray}
Denoting 
\begin{eqnarray}
&&Y=\cosh^2 \beta_3/4=  \nonumber \\
&&\cosh^2 \bar{\gamma}\cosh^2((\beta1+\beta_2)/4)) -\sinh^2\bar{\gamma}%
\cosh^2((\beta_1-\beta_2)/4)\;\;,  \nonumber
\end{eqnarray}
and inserting this back into (\ref{11}) we get 
\begin{equation}
{\rm Tr}\sqrt{\rho_1^{1/2}\rho_2\rho_1^{1/2}}=\frac{\sqrt{2\sinh \beta_1/4
\sinh \beta_2/4}} {\sqrt{\sqrt{Y}-1}}\;\;.  \label{15}
\end{equation}
All that remains is to express $\bar{\gamma}$ in terms of $%
(r_1,r_2,\theta_1, \theta_2)$. From (\ref{12.5}) we obtain 
\begin{equation}
\cosh^2 \bar{\gamma}=\cos^2 \Delta\theta\cosh^2\Delta r+
\sin^2\Delta\theta\cosh^2\Sigma r\;\;  \nonumber
\end{equation}
where $\Delta\theta=\theta_1-\theta_2$, $\Delta r=r_1-r_2$ and $\Sigma
r=r_1+r_2$. Defining $\beta_+=(\beta_1+\beta_2)/4$ and $\beta_-=(\beta_1-%
\beta_2)/4$ we finally get 
\begin{eqnarray}
Y&=&\cosh^2\beta_3/4  \nonumber \\
&=& \cos^2\Delta\theta\left[ \cosh^2\Delta r\cosh^2\beta_+-\sinh^2\Delta
r\cosh^2\beta_-\right]  \nonumber \\
&+& \sin^2\Delta\theta\left[ \cosh^2\Sigma r\cosh^2\beta_+-\sinh^2\Sigma
r\cosh^2\beta_-\right] \;\;,  \nonumber
\end{eqnarray}
and 
\begin{equation}
{\rm Tr}\sqrt{\rho_1^{1/2}\rho_2\rho_1^{1/2}}=\frac {2\sinh\beta_1/4\sinh%
\beta_2/4}{\sqrt{\sqrt{Y}-1}}\;\;,  \label{16}
\end{equation}
For the case $\Delta \theta=\Delta r=0$, ie. only a change in temperature, (%
\ref{16}) gives 
\begin{equation}
D^2_B(\rho(\beta_1),\rho(\beta_2))=2\left[1- \frac{\sqrt{\sinh\beta_1/4\sinh%
\beta_2/4}}{\sinh\displaystyle ( \frac{\beta_1+\beta_2}{8})}\right]\;\;.
\label{17}
\end{equation}

Equation (\ref{17}) gives the Bures distance between two thermal states with
temperatures proportional to $1/\beta_1$ and $1/\beta_2$. The distance
function (\ref{17}) (or more generally (\ref{1})) is a proper distance
function on the space of states. However, the resulting form (\ref{17}) is
clearly not a distance function arising from a local metric structure
defined on the parameter space. The restriction of the Bures distance to
pure states, the Fubini-Study distance, is derivable from a local metric ie. 
$D_{FS}^2=2(1-|\langle\psi_1|\psi_2\rangle|^2)= 2\cos\Delta\theta$ where $%
\Delta\theta$ is the angle between the two Hilbert space vectors $%
|\psi_1\rangle$, $|\psi_2\rangle$. From Uhlmann's derivation of the Bures
distance as the minimum Fubini-Study distance between the purifications of $%
\sqrt{\rho_1}$ and $\sqrt{\rho_2}$ in the larger Hilbert-Schmidt space we
see that the Bures distance arises from a metric structure in this larger
Hilbert-Schmidt space. To derive this metric we can proceed in two ways. We
can use standard perturbation analysis to evaluate $ds^2_B\equiv
D_B^2(\rho,\rho+\delta\rho)$. This was done in \cite{HUBNER} with the result 
\begin{equation}
\delta D^2_B=ds^2_B=\frac{1}{2}\sum_{i\neq j} \frac{\langle
u_i|\delta\rho|u_i\rangle\langle u_j|\delta\rho|u_i\rangle} {P_i+P_j}\;\;,
\label{20}
\end{equation}
where $|u_i\rangle$ are the eigenstates of $\rho$. This metric also appears
in \cite{BRAUNSTEINCAVES} and is known there as the Statistical Distance
metric. The Bures or Statistical Distance metric has been mostly studied in
the case of pure states \cite{INTEREST} while for impure states only a few
finite dimensional examples have been examined \cite{HUBNER1,HUBNER,DITTMANN}%
. Since we know the orthogonal eigenstates of $\rho$ we could compute (\ref
{20}) explicitly. This is simply done in the case $\Delta r=\Delta\theta=0$
but becomes quite tedious otherwise. Instead we note that 
\begin{eqnarray}
&&ds^2_B=g_{\alpha\beta}dx^\alpha dx^\beta=  \nonumber \\
&&\left.\frac{1}{2}\frac{d^2}{dt^2}D_B(\rho(\beta,r,\theta),\,
\rho(\beta+t\delta\beta,r+t\delta r,\theta+t\delta\theta))\right|_{t=0} \;\;.
\label{21}
\end{eqnarray}
Using (\ref{16}), (\ref{1}) and (\ref{21}) one can eventually obtain 
\begin{eqnarray}
ds^2_B&&=ds^2_{SD}  \nonumber \\
&&=\frac{1}{2}\left[ 1+{\rm sech} \beta/2\right](dr^2+\sinh^2 (2r) d\theta^2)
\nonumber \\
&&\mbox{  }+\frac{1}{64\sinh^2 \beta/4}d\beta^2\;\;,  \label{22}
\end{eqnarray}
where the subscripts indicate Bures or Statistical Distance. The metric may
be simplified somewhat by defining $\exp (-2u)=\tanh \beta/8$ to give 
\begin{eqnarray}
ds^2_B&=&ds^2_{SD}  \nonumber \\
&=&\frac{1}{2}\left[1+\tanh^2 u\right](dr^2+\sinh^2 (2r) d\theta^2)
+du^2\;\;.  \label{23}
\end{eqnarray}
As a check we have directly computed the $d\beta^2$ contribution from (\ref
{20}) while for pure states ($\beta\rightarrow\infty$) the metric reduces to
the known form \cite{KNOWN}. Equipped with the metric (\ref{22}), one can
compute geometrical quantities such as the scalar curvature 
\begin{equation}
R_{SD}=-\frac{8\left( \cosh^2\beta/4+12\sinh^4\beta/4\right)} {\cosh^2\beta/2%
}\;\;.  \label{24}
\end{equation}
It is interesting to note that the scalar curvature is independent of the
``unitary" parameters $r$ and $\theta$ and only depends on the ``nonunitary"
parameter $\beta$. This may be understood from the work of Dittmann \cite
{DITTMANN}. Dittmann shows that in finite dimensions the geometry of the
parameter space is locally isometric to $S^{n-1}\times U(n)/T^n$. The
homogeneous submanifold $U(n)/T^n$, which, in our case, is parametrised by $%
r(\beta)$ and $\theta(\beta)$, has a constant curvature depending only on $%
\beta$. Thus, in the general case, the curvature should only be a function
of the invariants $\beta_i$ \cite{LENDI}  of the state $\rho$.

\section{Distinguishability Measure}

A physical significance can be attributed to the curvature $R_{SD}$ (\ref{24}%
). From a statistical inference viewpoint the Statistical Distance can be
understood as a measure of how well one can, in principle, determine the
parameters describing $\rho$ through $N$ arbitrary generalised measurements.
Fro more on this viewpoint see \cite{BRAUNSTEINCAVES,MILBURN}.  The error $%
\delta X$ in estimating the parameter $X$ by analysing the data obtained
from $N$ copies of $\rho (X)$ is bounded by 
\begin{equation}
N\langle(\delta X)^2\rangle\left(\frac{ds}{dX}\right)^2\geq 1\;\;,
\label{25}
\end{equation}
where $ds/dX$ is the rate of change of statistical Distance with respect to
the parameter $X$ for a single copy of $\rho(X)$. Thus, if two states are
separated by a Statistical Distance of $ds$ then one must perform at least $%
N\geq 1/ds^2$ measurements on identically prepared copies to distinguish
between the two. To estimate a parameter seperation between two states one
can make use of (\ref{25}).

However, to go beyond distinguishing between two states and to provide a
complete estimation of a state given a reference state one can argue that
the accuracy of estimation of the {\em complete} state  should be
independant of the particular parametrisation used. The most natural
geometric quantity which is parametrisation (coordinate) independant is the
scalar curvature of the Statistical Distance. Following this argument, we
can interpret the curvature $R_{SD}$ as the degree of local
distinguishability of the complete state i.e. if $R_{SD}$ is large then few
measurements are needed to estimate the parameters of a neighboring state
while if $R_{SD}$ is small, many measurements will be required to estimate
the parameters of a neighboring state. We plot the behavior behavior $%
R_{SD}(\beta)$ in Figure 1. From the form of $R_{SD}$ (\ref{24}) we see that
the degree of distinguishability diverges as $\beta\rightarrow\infty$.
Similarly, from (\ref{22}) we find that the integrated Statistical Distance
between a pure state and {\it any} impure state diverges. This feature was
also seen in a similar calculation by Braunstein and Milburn \cite{MILBURN}.
As was explained there, the degree of distinguishability, $R_{SD}$, diverges
at the pure state boundary because only one measurement is needed to
differentiate between a pure state with $\beta=\infty$ and  an impure state
with {\em any} $\beta<\infty$. This principle was the basis of the one-shot
clock in \cite{MILBURN} where the parameter to be estimated was time.  From (%
\ref{24}) we can also identify local maxima (at $\beta=0$) and minima (at $%
\beta=4\cosh^{-1}5/\sqrt{22}$). The ultimate accuracy of parameter
estimation at various values of $\beta$ is reflected in the behavior of the
degree of distinguishability $R$. This measure can serve as a guideline for
optimal operating regimes in quantum non-demolition measurements. By
targeting the measurements to operate in those regions of high
distinguishability one can obtain information about the quantum system in
the least number of measurements.

\section{Conclusion}

In this paper we have examined the geometry of the quantum state. After
reviewing the previous work we showed how one can extend the definition of
the quantum metric to deal with impure quantum states. The resulting metric
is known as the Bures metric. To compute the geometry of a quantum state
using this Bures metric necessitates the computation of the square-root of a
density matrix. Up till now this has only been done for very simple quantum
systems. Using a Shur decomposition of $\rho$ we have calculated the Bures
distance and associated Riemaniann metric between two squeezed thermal
states. The Riemaniann manifold is not of constant curvature as was
suggested by spin-1/2 calculations. It was argued that the scalar curvature
of the Riemaniann manifold can serve as a measure of the ultimate accuracy
in determining the parameters defining a quantum state. The method used here
can be generalised to more complicated quantum states if a Schur
factorisation can be found. More generally, computable methods for solving
the finite or infinite dimensional Lyapunov equations need to be
investigated before one can understand the geometric structures of high
dimensonal impure quantum states.

\section*{Acknowledgements}

The author thanks C. A. Hurst and J. McCarthy for useful discussions and M.
Nielsen for his comments on the manuscript.

\begin{description}
\item  {\bf Figure 1}: Graph of the curvature of the Statistical Distance
metric $R_{SD}$ vs. inverse temperature $\beta $ in equation (\ref{24}).
\end{description}

\end{document}